\documentclass[12pt]{iopart}

\usepackage{amssymb}
\usepackage{graphicx}
\usepackage{multirow}
\usepackage{booktabs,siunitx}
\usepackage{booktabs}

\begin{document}

\title[]{Magnetic properties of disordered polycrystalline bulk Sm$ _{2} $NiMnO$ _{6} $ double perovskite}

\author{
S. Majumder$^{a}$,
\
M. Tripathi$^{a}$,
\
P. Rajput$^{b}$,
\
S. N. Jha$^{b}$,
\
R. J. Choudhary$^{a}$,
and
D. M. Phase$^{a}$
\\
$^{a}$UGC DAE Consortium for Scientific Research, Indore 452001, India\\
$^{b}$Beamline Development and Application Section, Bhabha Atomic Research Centre, Mumbai 400085, India\\}
\ead{$^{*}$ram@csr.res.in}

\begin{abstract}
The structural, electronic and magnetic properties of anti-site disordered Sm$ _{2} $NiMnO$ _{6} $ double perovskite has been studied. RE$_{2}$NiMnO$_{6}$ (RE: rare-earth) ordered double perovskite is commonly believed to show two distinct magnetic phase transitions viz, paramagnetic to ferromagnetic (FM) transition at T = T$ _{C} $ due to Ni-O-Mn super exchange interaction and another transition at T = T$ _{d} $ due to coupling of RE spins with Ni-Mn network. In our present study, we have observed that the presence of intrinsic B-site disorder results in an additional antiferromagnetic (AFM) coupling, mediated via Ni-O-Ni and Mn-O-Mn local bond pairs. As a consequence, the magnetic behavior of SNMO comprises of co-existing FM-AFM phases, which are respectively governed by the anti-site ordered and disordered structures. Field dependent inverted cusp like trend in M(T) and two step reversible loop behavior in M(H) measurements indicate the presence of competing FM-AFM phases over a wide range of temperature values (T$ _{d} < $ T $ < $ T$ _{C} $).
\end{abstract}

\section{Introduction}
The multifunctionality of double perovskite (DP) oxide A$_{2}$B$'$B$''$O$_{6}$ systems is significantly altered by the mis-locations of transition metal ions (B$'$ and B$''$) from ideal alternating site occupancy, a phenomenon commonly known as anti-site disorders \cite{HZGuo2008, MPSingh2009, JBenitez2011, VNSingh2011, PSanyal2017}. In perfectly ordered DP structure, the B$'$, B$''$ transition metals are arranged in ideal rock salt fashion kinetically favored by significant differences in ionic radius and formal charge states between B$'$, B$''$ cations \cite{MTAnderson1993}. The presence of two transition metals and consequently, a variety of electronic and magnetic interactions lead to a complicated phase diagram in the DP structures. The electronic and magnetic ground states of DP systems are highly sensitive to the choice of B$'$, B$''$ cations, for instance; the ground state of Sr$ _{2} $FeMoO$ _{6} $ is half metallic ferromagnetic, Sr$ _{2} $FeReO$ _{6} $ is half metallic ferrimagnet, Sr$ _{2} $FeWO$ _{6} $ is antiferromagnetic insulator whereas La$ _{2} $FeCoO$ _{6} $ is ferromagnetic semiconductor etc. \cite{KIKobayashi1998, KIKobayashi1999, HKawanakaet1999, HRFuh2015}. The presence of anti-site disorder introduces further complexity in the phase diagram of DP structures \cite{SMajumder2022prbb, SMajumder2022prbt, SMajumder2022jpcm}. In the present work, we aim to investigate the structural, magnetic and electronic properties of so far un-explored B-site disordered Sm$_{2}$NiMnO$_{6}$ (SNMO) double perovskite oxide. Interestingly, we have observed that the magnetic behavior of SNMO comprise of co-existing FM and AFM phases in a broad temperature range ($ \Delta T \sim $ 100 K in presence of $ \mu_{0}H $ = 0.01 T). 

\section{Methodology}
Polycrystalline sample of SNMO was synthesized by solid state reaction route. Stochiometric amount of starting materials Sm$ _{2} $O$ _{3} $ (99.9 \%), NiO (99.99 \%), MnO$ _{2} $ (99.9 \%), were grinned to form a homogeneous mixture and then calcined at 900$ ^{o} C $ for 24 h. The phase and structural charecterization of the sample, at room temperature, were done by $ 2\theta $ scans using Cu K$ \alpha $ ($ \lambda$=1.54 {\AA}) X-ray diffraction (Bruker D2 Phaser Desktop Diffractometer). Chemical valence state of the elements present in the sample, were investigated by X-ray photo electron spectroscopy (XPS) experiments using Al K$ _{\alpha} $ ($ h\nu $ = 1486.7 eV) lab-source and Omicron energy analyzer (EA-125) at angle integrated photoemission spectroscopy (AIPES) beamline (Indus-1, BL 2, RRCAT, Indore, India) and X-ray absorption near edge spectroscopy (XANES) measurements in fluorescence mode using hard X-ray synchrotron radiation (Indus-2, BL 9, RRCAT, Indore, India). The charging effect corrections in XPS, are accounted by measuring C 1\textit{s} core level spectra. All of these spectra are recorded at 300 K. XPS spectra are deconvoluted by fitting with combined Lorentzian-Gaussian function and Shirley background. The estimated experimental resolution for XPS and XANES measurements across the measured energy range are about 0.6 eV and 0.25 eV, respectively. Magnetization measurements were performed using MPMS 7 T SQUID-VSM instrument (Quantum Design Inc., USA). Before recording the magnetometric data, every time standard \textit{de}-\textit{Gaussing} protocol was followed to eliminate the trapped magnetic field of the superconducting coil inside the magnetometer, and the sample was heated well above its magnetic transition temperatures to erase the prior magnetic history (if any).

\begin{figure}[t]
\centering
\includegraphics[angle=0,width=0.95\textwidth]{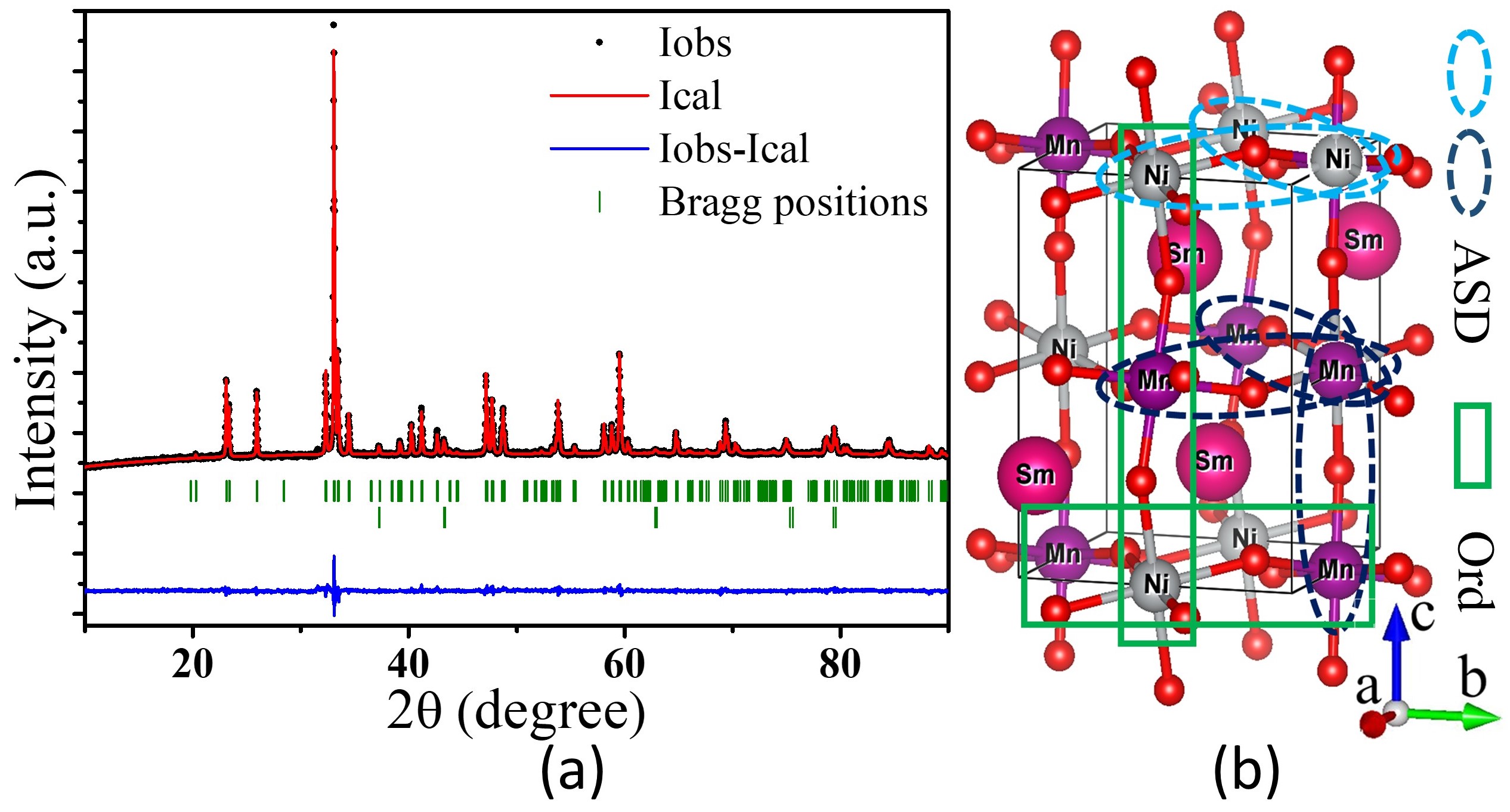}
\caption{(a): Reitveld analysis of X-ray diffraction $2\theta $ profile showing observed, calculated and difference pattern along with Bragg positions for Sm2NiMnO6 \textit{P2$_{1}$/n} and NiO \textit{Fm -3m} structures. (b): Crystal structure of SNMO showing Ni/Mn anti-site disordered (ASD) and ordered bonds.}\label{XRD structure}
\end{figure}

\section{Results and Discussions}
\subsection{Structural properties}
The powder X-ray diffractogram of SNMO sample recorded at room temperature can be matched with Rietveld generated model patterns for both monoclinic \textit{P2$_{1}$/n} (SG 14) and orthorhombic \textit{Pbnm} (SG 62) structures, as shown in Fig. \ref{XRD structure}(a). This is because, $ \surd 2a_{p} \times \surd 2a_{p} \times 2a_{p} $ (where average $a_{p}$ stands for pseudo-cubic) unit cell is subject to both \textit{P2$_{1}$/n} and \textit{Pbnm} space groups, depending on B-site ordered or disordered structures, respectively. However, in present case it is not easy to distinguish between \textit{P2$_{1}$/n} and \textit{Pbnm} symmetries by X-ray diffraction because, (i) mean scattering factor for X-ray from Ni and Mn are almost equal, (ii) small difference in $ \beta $ for \textit{P2$_{1}$/n} and \textit{Pbnm} \cite{SMajumder2022prbb}. In our analysis we have considered the \textit{P2$_{1}$/n} structural model for SNMO. The reliability factors for the refinement, $ \chi^{2} = 2.23 $, $ R_{p} = 16.1 $, $ R_{wp} = 10.9 $, $ R_{exp} = 7.28 $ indicate a reasonable agreement between experimentally observed data and simulated pattern.A schematic representation of the refined crystal structure consisting anti-site disorder and ordered bonds is illustrated in Fig. \ref{XRD structure}(b). The cell parameters obtained from refinement are, $ a=5.35570(8) \AA $, $ b=5.53895(9) \AA $, $ c=7.61694(12) \AA $, and $ \beta=90.03^{o}(2) $. Traces of residual precursor oxide is detected as secondary phase (7.27(2)$\%$), which is quite common in rare-earth based double perovskite synthesis \cite{HNhalil2015, CShi2011, HSNair2011, WZYang2012} and is identified as NiO (\textit{Fm -3m}, SG 225). 

\begin{figure}[t]
\centering
\includegraphics[angle=0,width=0.95\textwidth]{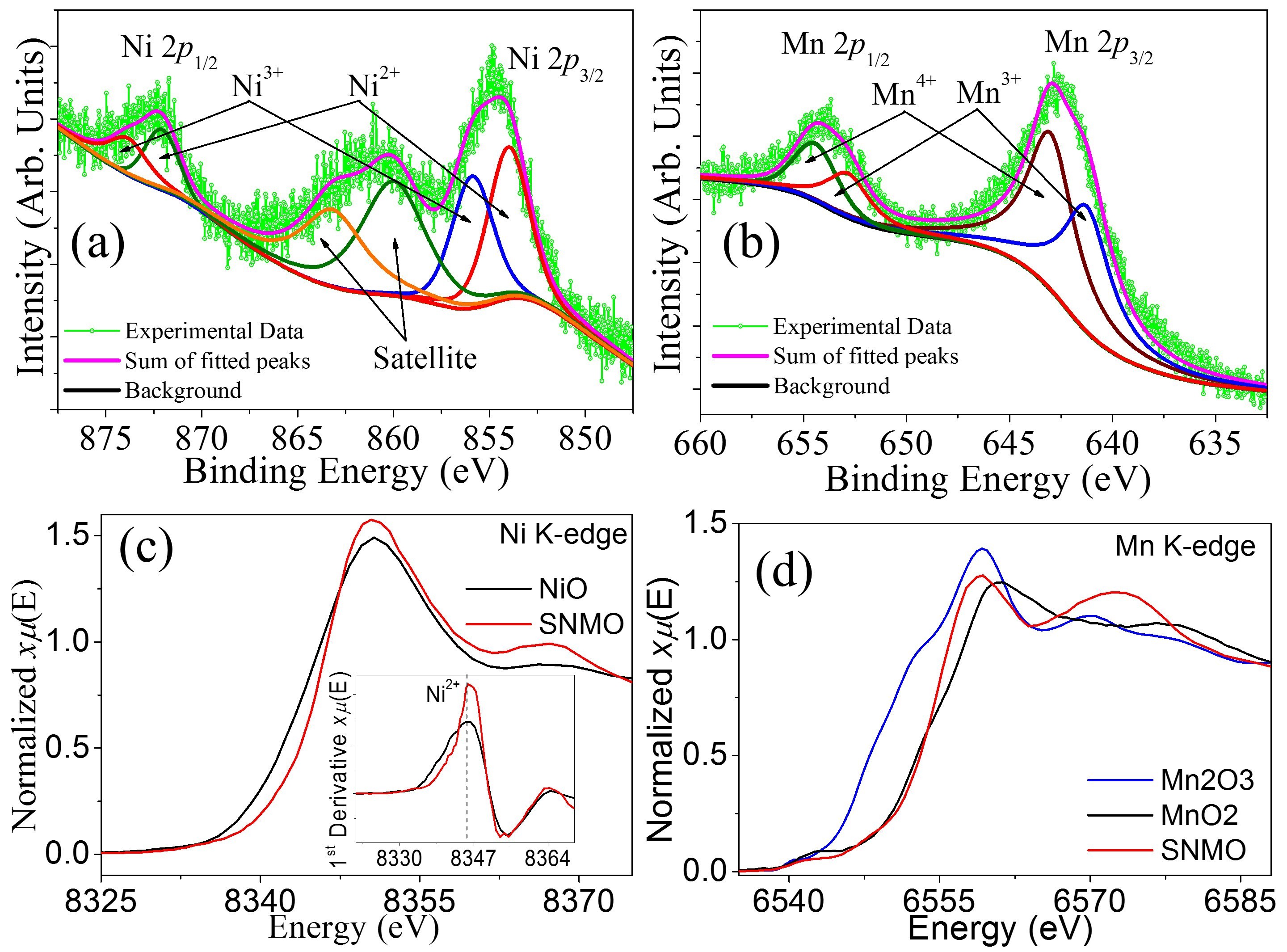}
\caption{X-ray photo electron spectroscopy results of (a): Ni, and (b): Mn 2\textit{p} core level spectra showing presence of Ni$ ^{2+/3+} $ and Mn$ ^{4+/3+} $ species. X-ray absorption near edge spectroscopy results of Ni and Mn \textit{K}-edge spectra suggesting mixed valency for both ions.}\label{SNMO bulk data E}
\end{figure}

\subsection{Chemical valence state}
In order to probe the valence state of constituent elements present in the sample, XPS measurements are carried out. The core level XPS spectra of Sm 3\textit{d} (not shown here) suggests that Sm is in 3+ valence state. Ni and Mn core level spectra splitted into 2\textit{p}$ _{1/2} $ and 2\textit{p}$ _{3/2} $ due to spin-orbit spliting are shown in Fig. \ref{SNMO bulk data E}(a, b). The asymmetry and broadening of spectral characters observed in XPS for both Ni and Mn, suggest the presence of more than one valence features. Thus, Ni 2\textit{p} XPS spectra can be deconvoluted by fitting it with peaks contribution from both Ni$ ^{2+} $ and Ni$ ^{3+} $ features with peaks centered at 854.0 eV (Ni$^{2+}$ 2\textit{p}$ _{3/2} $), 872.1 eV (Ni$^{2+}$ 2\textit{p}$ _{1/2} $), 855.9 eV (Ni$^{3+}$ 2\textit{p}$ _{3/2} $) and 873.8 eV (Ni$^{3+}$ 2\textit{p}$ _{1/2} $) \cite{JWang2016}. Similarly, deconvolution of Mn 2\textit{p} XPS spectra reveals the presence of both Mn$ ^{4+} $ and Mn$ ^{3+} $ states centered at binding energies 643.1 eV (Mn$^{4+}$ 2\textit{p}$ _{3/2} $), 654.4 eV (Mn$^{4+}$ 2\textit{p}$ _{1/2} $) \cite{YUmezawa1920}, 641.2 eV (Mn$^{3+}$ 2\textit{p}$ _{3/2} $) \cite{MOku1975} and 652.8 eV (Mn$^{3+}$ 2\textit{p}$ _{1/2} $). To further investigate whether these mixed valence nature are only surface properties (because XPS is only surface sensitive) or similarly distributed in depth of the sample also, we have used the large penetration depth of hard X-ray and measured the XANES spectra across Ni and Mn K-edge for SNMO and standard samples NiO, MnO$ _{2} $, Mn$ _{2} $O$ _{3} $ (shown in Fig. \ref{SNMO bulk data E}(c, d)). With increasing valency of transition metal (TM) inos the band edge shifts to higher energies \cite{AHdeVries2003}. Comparing the TM \textit{K} edge XANES spectra for SNMO with standard samples, it can be inferred that, (i) Ni has valency greater than 2+, (ii) Mn has valency in between 3+ and 4+, which are consistent with the XPS results. Bond valence sum (BVS) method is also used for valency calculations. Under this formalism the effective valency between a cation and neighboring anion is defined as $ V_{i} = exp[(d_{0}-d_{i})/B] $, where $ d_{0} $ is the empirically determined bond valence parameter for the cation anion pair, $ d_{i} $ is the bond distance of the cation anion bond and B = 0.37 is a universal constant \cite{OChmaissem2001}. The calculated valencies of Ni and Mn species using SpuDs \cite{MWLufaso2006, MWLufaso2001} software following BVS method (Ni$_{BVS}$ valency: 2.69, Mn$_{BVS}$ valency: 3.33) also indicates mixed valence nature for both Ni$ ^{2+/3+} $ and Mn$ ^{4+/3+} $ ions. In DP structure, charge disproportion between TM sites like $ Ni^{2+}+Mn^{4+} \longrightarrow Ni^{3+}+Mn^{3+} $ is favored, which causes mixed valency in cations \cite{MPSingh2009, HGuo2006, HZGuo2008, NSRogado2005}. 

\begin{figure}[t!]
\centering
\includegraphics[angle=0,width=1.0\textwidth]{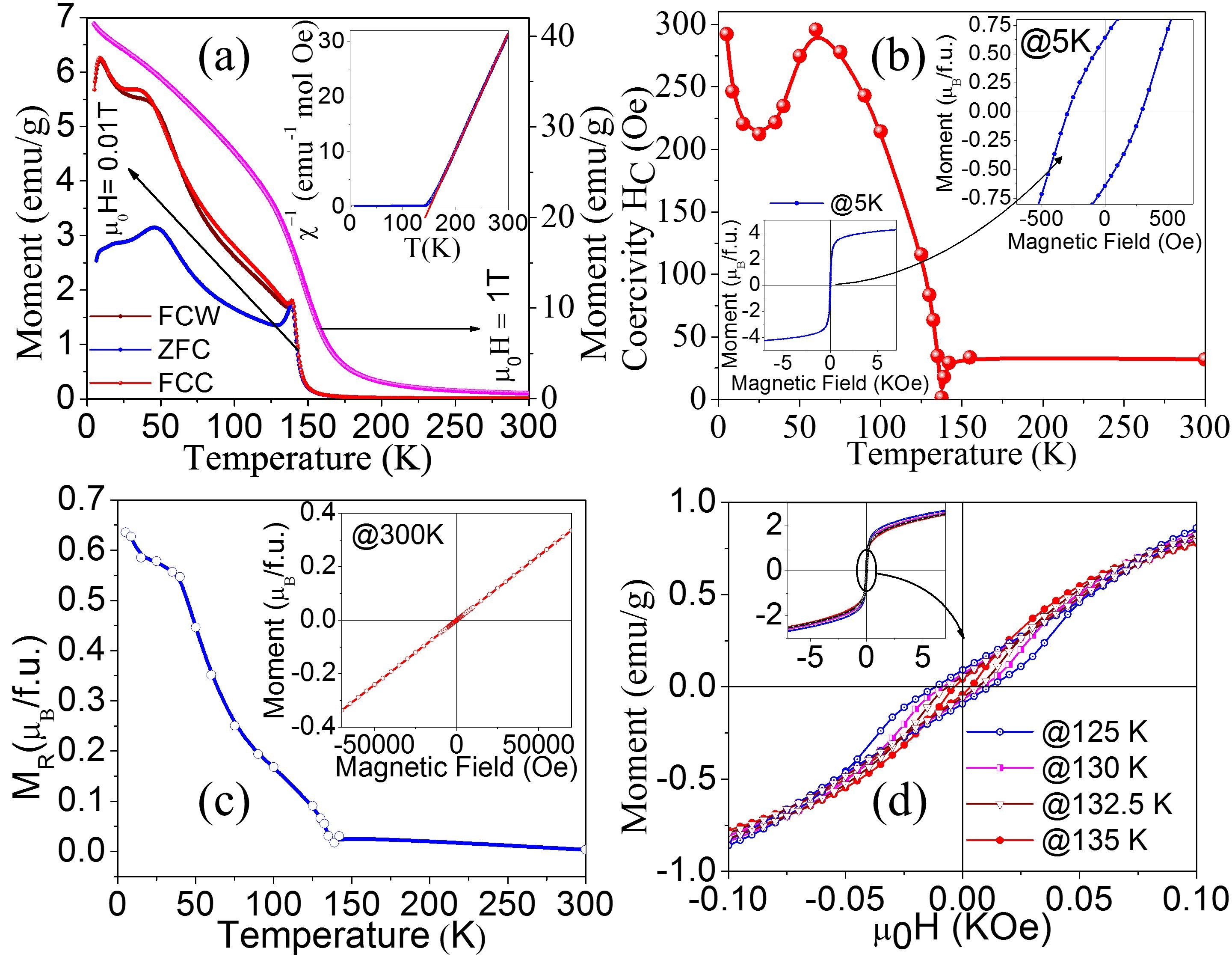}
\caption{(a): Temperature dependence of magnetization measured in presence of $ \mu_{0}H $ = 0.01, 1 T. Inset shows the CW fitting of inverse susceptibility. (b, c): Thermal evolution of coercivity and remanence magnetization. Insets show magnetic hysteresis isotherms at T = 5 K and 300 K. (d): Two-step reversible M(H) loops measure in different temparetures within $ T_{d} < T < T_{C} $.}\label{SNMO bulk data M}
\end{figure}

\subsection{Magnetic properties}
To explore the anti-site disorder driven modifications in magnetic properties of SNMO we have measured dc bulk magnetization as a function of temperature in presence of different applied magnetic fields and as a function of applied magnetic fields in isothermal conditions, as shown in Fig. \ref{SNMO bulk data M}. Temperature dependent magnetization measured in $ \mu_{0}H $ = 0.01 T following zero field cooled warming (ZFC), field cooled cooling (FCC) and field cooled warming (FCW) protocols show two distinct transitions at $ T_{C} \sim $ 150 K and $ T_{r} \sim $ 46 K. The high temperature linear part of reciprocal susceptibility measured in ZFC mode in presence of $ \mu_{0}H $ = 0.01 T (inset of Fig. \ref{SNMO bulk data M}(a)) is fitted to classical Curie-Weiss (CW) law with Curie constant C = 4.75 emu mol$ ^{-1} $ K$ ^{-1} $ and paramagnetic (PM) CW temperature $ \theta \sim $ 150.3 K. The positive value of $ \theta $ suggests ferromagnetic interaction is dominating in the system. The effective PM moment calculated from CW fitting is found to be $ \mu_{expt} $ = 6.16 $ \mu_{B} $. The spin only theoretical effective magnetic moments for different spin configurations defined as, $ \mu_{theo} = \sqrt{ 2\times\mu_{Sm}^{2}+\mu_{Ni}^{2}+\mu_{Mn}^{2} } $, where $ \mu_{Sm} = g \sqrt {J(J+1)}  $ with $g \simeq $ 0.2857 and $ \mu_{{Ni/Mn}} = g \sqrt {S(S+1)}$ with $g \simeq $ 2.00, yield values \cite{SMajumder2022prbb}: 4.94 $ \mu_{B} $ for [Sm$^{3+}$ (J=5/2), Ni$ ^{2+}$ (S=1), Mn$ ^{4+}$ (S=3/2)], 6.36 $ \mu_{B} $ for [Sm$^{3+}$ (J=5/2), Ni$ ^{3+}_{HS}$ (S=3/2), Mn$ ^{3+}_{HS}$ (S=2)], and 5.33 $ \mu_{B} $ for [Sm$^{3+}$ (J=5/2), Ni$ ^{3+}_{LS}$ (S=1/2), Mn$ ^{3+}_{HS}$ (S=2)]. Therefore, $ \mu_{expt} $ value is observed to lie in between the calculated $ \mu_{theo} $ values for (Ni$ ^{2+} $ + Mn$ ^{4+} $) and mixed spin state of (Ni$ ^{3+} $ + Mn$ ^{3+} $). RE$ _{2} $NiMnO$ _{6} $ (RE: rare-earth) ordered double perovskite is believed to show two distinct magnetic phase transitions, one is PM-FM second order transition at temperature $ T_{C} $ due to Ni$ ^{2+} $-O-Mn$ ^{4+} $ (or Ni$ ^{3+} $-O-Mn$ ^{3+} $) superexchange interaction and another at temperature $ T_{d} $, due to coupling of RE spins with Ni-Mn network \cite{WZYang2012, JBenitez2011}. According to the Goodenough-Kanamori rules, the superexchange interactions between half-filled $ e_{g} $ and empty $ e_{g} $ orbitals, through Ni$ ^{2+}(e_{g}^{2}) $-O-Mn$ ^{4+}(e_{g}^{0}) $ is FM for 180$ ^{0} $ bond angle and it changes to antiferromagnetic (AFM) type when the angle becomes 90$ ^{0} $ \cite{JBGoodenough1976, JJKanamori1959, JBGoodenough1955}. We observe that the presence of intrinsic B-site disorder results in AFM coupling also, mediated by virtual charge transfer between half-filled to half-filled orbital via Ni$ ^{2+}(e_{g}^{2}) $-O-Ni$ ^{2+}(e_{g}^{2}) $ and Mn$ ^{4+}(t_{2g}^{3}) $-O-Mn$ ^{4+}(t_{2g}^{3}) $ pairs \cite{HZGuo2008}. In addition, the vibronic superexchange interactions between Jahn-teller active singly occupied twofold degenerate fluctuating $ e_{g} $ orbitals give rise three-dimensional FM via Ni$ ^{3+}(e_{g}^{1}) $-O-Mn$ ^{3+}(e_{g}^{1}) $, Ni$ ^{3+}(e_{g}^{1}) $-O-Ni$ ^{3+}(e_{g}^{1}) $ and Mn$ ^{3+}(e_{g}^{1}) $-O-Mn$ ^{3+}(e_{g}^{1}) $ \cite{HZGuo2008}. So the coexistence of FM phase along with AFM phase greatly affects the magnetic properties of SNMO system. $ (Ni/Mn)O_{6} $ octahedral tilting results in reduction of the average bond angle $ \angle (Ni/Mn-O-Ni/Mn) $ in SNMO causing a decrease in T$ _{C} $. Another possibility of reduction in T$ _{C} $ is may be due to cationic disorder mediated weakenning of FM interactions \cite{HZGuo2008, MPSingh2009}. We have observed an inverted cusp like trend, possibly due to cataonic disorder and thermal irreversibility in FCC and FCW cycles in the temperature range of $ T_{r} < T < T_{C} $. On application of higher magnetic fields, B-site disorder effect weakens as well as thermal hysteresis vanishes (Fig. \ref{SNMO bulk data M}(a)). The observed thermal hysteresis indicates towards the possibility of inhomogenity in magnetic response for cooling and warming paths due to either magnetically phase separated species \cite{HSNair2011}, or first order phase transition in coexisting FM-AFM phases or magnetic frustration originating from different exchange interaction paths between mixed valence ions \cite{Pradheesh2012}. The thermal evolution of coercivity (H$ _{C} $) and remanence magnetization (M$ _{R} $) shown in Fig. \ref{SNMO bulk data M}(b, c) suggest non-monotonic behavior. At T = 5 K the applied magnetic field $ \mu_{0}H $ = 7 T is not sufficient to saturate the magnetization curve (inset of Fig. \ref{SNMO bulk data M}(b)), whereas at T = 300 K M($ \mu_{0}H $) shows (inset of Fig. \ref{SNMO bulk data M}(c)) typical paramagnetic type straight line nature without saturation. The isothermal moment value obtained M$ _{S} $(at T=5 K, $ \mu_{0}H $ = 7 T) = 4.26 $ \mu_{B} $ is found very low as compared to highly cation ordered SNMO case (M$ _{S} $= 5.56 $ \mu_{B} $) \cite{SMajumder2022prbb}. This indicate a substantial presence of anti-site disorder in the studied SNMO sample. Magnetic hysteresis loop isotherms measured at different temperature values in $ T_{r} < T < T_{C} $ region exhibit two step reversibility loop behavior, which further confirms the presence of competing FM-AFM phases (Fig. \ref{SNMO bulk data M}(d)). The NiO secondary phase may also show long range AFM ordering, if they form phase segregated clusters in the sample, but assuming their random distribution, we can argue that the NiO contribution in observed magnetic behavior is nominal. Therefore, in studied SNMO system the magnetization properties are governed by phase fractions of cation ordered and disordered structures.

\section{Conclusion}
In summary, we have studied the structural, electronic and magnetic properties of anti-site disordered SNMO. Structural characterization reveals that SNMO crystalizes in monoclinic \textit{P2$_{1}$/n} structure. The observed mixed valence nature of both Ni and Mn species confirms the $ Ni^{2+}+Mn^{4+} \longrightarrow Ni^{3+}+Mn^{3+} $ charge disproportion in the system. Mislocation of B-site cation results in additional Ni-O-Ni and Mn-O-Mn AFM superexchange interactions along with Ni-O-Mn FM coupling originating due to ordered structure. Consequently, the magnetic behavior of SNMO comprise of co-existing FM and AFM phases in a broad temperature range ($ \Delta T \sim $ 100 K in presence of $ \mu_{0}H $ = 0.01 T). Competing nature of these two phases left its imprints in magnetization behavior, such as: field dependency of (i) inverted cusp like trend, (ii) thermal irreversibility in cooling and warming paths of M(T) cycle, (iii) two step reversible loop behavior in M(H) isotherms, (iv) non-monotonic nature is observed in temperature dependence of  $ H_{C} $ and $ M_{R} $. The present study will in general, help to understand the alteration in functional properties of DP systems by introducing anti-site disorders in the host matrix.

\section*{Acknowledgments}
Thanks to Indus Synchrotron RRCAT, India for giving access to experimental facilities. S.M. thanks Mr. Avinash Wadikar (CSR, India), and Mr. Sharad Karwal (CSR) for their technical help in XPS measurements performed at RRCAT.


\end{document}